\documentclass{article}
\usepackage{spconf,amsmath,graphicx,hyperref, multirow, booktabs, array}
\usepackage{enumitem}
\usepackage{amssymb}

\title{
ADVANCED MODELING OF INTERLANGUAGE SPEECH INTELLIGIBILITY BENEFIT \\
WITH L1-L2 MULTI-TASK LEARNING USING DIFFERENTIABLE K-MEANS \\
FOR ACCENT-ROBUST DISCRETE TOKEN-BASED ASR \\}

\name{Kentaro Onda$^{\star \dagger}$ \quad 
Satoru Fukayama$^{\dagger}$ \quad 
Daisuke Saito$^{\star}$ \quad
Nobuaki Minematsu$^{\star}$}
  
  \address{$^{\star}$ The University of Tokyo \quad
      $^{\dagger}$National Institute of Advanced Industrial Science and Technology (AIST)}

\begin{document}
\ninept
\maketitle
\begin{abstract}
Building ASR systems robust to foreign-accented speech is an important challenge in today’s globalized world. 
A prior study explored the way to enhance the performance of phonetic token-based ASR on accented speech by reproducing the phenomenon known as interlanguage speech intelligibility benefit (ISIB), where foreign-accented speech is more intelligible to listeners sharing the speaker’s native language than to native listeners.
ISIB was technically implemented by using the speaker’s L1 to learn k-means cluster centroids in an SSL feature space to obtain phonetic tokens.
In this study, we propose a more advanced modeling of ISIB. By employing differentiable k-means and optimizing the entire module for both L1 and L2 ASR, the proposed method outperformed the baselines, both when using only native speech and when additionally incorporating a limited amount of accented speech. Notably, in the latter scenario, our method achieved approximately a 20\% relative improvement in recognition accuracy.

\end{abstract}
\section{Introduction}
\label{sec:intro}
Recent advances in automatic speech recognition (ASR) have achieved recognition accuracy comparable to that of humans \cite{asr_parity, asr_survey}. However, certain types of speech, such as noisy or spontaneous speech, remain still challenging in practical applications. Foreign accent is also one of these challenges. It is well known that recognition accuracy tends to degrade when recognition is performed on accented
speech produced by non-native speakers \cite{racial}. In today’s globalized world, where people often communicate in foreign languages with diverse accents \cite{munro1995foreign, murphy2014intelligible, levis2020revisiting}, the demand for accent-robust ASR has been even increasing.

To achieve accent-robust ASR, a variety of approaches, such as data augmentation and accent adaptation, have been proposed \cite{fukuda18_interspeech,Klumpp2023SyntheticCD,prabhu-etal-2023-accented,prabhu24b_interspeech}. However, these approaches typically assume the availability of foreign-accented speech as training data. Most existing foreign-accented speech corpora consist of “X-accented English,” while corpora of foreign-accented speech in other languages are extremely scarce \cite{hinsvark2021accentedspeechrecognitionsurvey}. Given this limitation, it seems natural to explore ways of leveraging native speech data, of both the speaker's native language (L1) and the language spoken (L2), in a sophisticated manner. Since native speech samples can be collected more easily, such approaches may be effectively applied to a wider range of ``X-accented Y" speech.

In \cite{onda25_interspeech}, a method was proposed to improve ASR robustness to foreign-accented speech using only native speech data by reproducing a phenomenon observed in human listeners called interlanguage speech intelligibility benefit (ISIB). ISIB refers to the tendency that foreign-accented speech is more intelligible when the speaker and the listener share the same native language (L1), compared to when the listener is a native speaker of the language spoken (L2) \cite{bent2003interlanguage,harding2012accent,XIE2013369}. For example, Japanese-accented English is more easily understood by Japanese listeners than by American listeners. In \cite{onda25_interspeech}, the simulation of ISIB was performed within the framework of discrete token-based ASR, which has been actively studied in recent years \cite{chang23b_interspeech, chang2024exploring, yang2024towards, chang24b_interspeech, mousavi24_interspeech}. 
Here, the discrete tokens used are phonetic tokens (also referred to as semantic tokens) \cite{lakhotiaetal2021generative, guo2025recentadvancesdiscretespeech, mousavi2025discreteaudiotokenssurvey}, which are obtained by applying k-means clustering to the outputs of a self-supervised learning (SSL) model \cite{hubert, sslreview}. This study was conducted based on the hypothesis that these phonetic tokens can simulate the categorical perception of speech by native speakers of the language used for k-means training. \cite{onda24_interspeech, onda25b_interspeech}.
Experimental results confirmed that ISIB actually occurs in discrete token-based ASR. For example, when recognizing Japanese-accented English, using phonetic tokens learned from Japanese (L1) as intermediate representations, rather than those learned from English (L2), improved recognition accuracy.

However, this approach appears somewhat naive as a simulation of ISIB. This is because the method relied solely on the listener’s native language to simulate his/her speech perception, using k-means clustering to obtain phonetic tokens.
In reality, however, it depends on listeners' language background how their speech perception is biased toward their L1, and the degree of this bias is likely to change during the process of learning a new language \cite{flege1995second, stolten2014effects}. Listeners who share the same native language as the speaker, who are assumed in the framework of ISIB to better understand accented speech, are of course also learners of the L2 themselves, and their speech perception should be influenced by both their L1 and L2, rather than by the L1 alone.
Thus, to reproduce ISIB in a more realistic manner, the discrete tokens should be optimized using both the listener’s (and the speaker's) L1 and the language spoken (L2). In conventional discrete token-based ASR, a two-step approach has been widely used: tokens are first learned, and then downstream ASR is trained using these tokens. In \cite{onda25_interspeech} as well, k-means cluster centroids were first learned from the listener’s L1, and then the ASR was trained on native speech of the target language spoken (L2), which was converted into tokens using those centroids. 

In \cite{ghorbani18_interspeech}, it was reported that multi-task learning, which optimizes a shared encoder for ASRs of both the speaker’s L1 and the target language (L2), can improve recognition performance for non-native speech. In this study, we introduce this multi-task learning approach into discrete token-based ASR using the recently proposed differentiable k-means \cite{onda25c_interspeech}, aiming to reproduce ISIB more validly and enhancing the ASR performance.

The key strengths of our method are summarized as follows:
\begin{itemize}[left=0pt]
    \item \textbf{Advanced modeling of ISIB:} By employing differentiable k-means along with multi-task learning for L1 and L2 ASR, joint optimization of the entire system was made possible. This addresses a limitation of conventional ISIB modeling, where the tokenization, corresponding to the simulation of categorical speech perception by humans, is trained solely on L1. The proposed method enables a more valid reproduction of speech perception by non-native listeners, and improves recognition accuracy for foreign-accented speech.

    \item \textbf{Leveraging abundant native speech data available:} Our method primarily leverages abundant native speech data available to enhance the robustness of ASR to non-native speech. 
    We achieved improved recognition accuracy in two scenarios: 1) \textbf{native-only}, where no additional accented speech is used, and 2) \textbf{accent-adapted}, where only a small amount (2–5h) of accented speech is additionally incorporated.

    \item \textbf{A large performance gain:} In the second \textbf{accent-adapted} scenario, where abundant native speech data for the two languages (L1 \& L2) and a limited amount of accented speech data are used, our method achieved an approximately 20\% relative reduction in word error rate (WER) compared to the vanilla k-means baseline.
\end{itemize}

\section{Advanced modeling of interlanguage speech intelligibility benefit}
\label{sec:method}
\begin{table*}[tb]
	\centering
	\caption{Recognition performance on out-of-domain speech from ERJ (native and Japanese-accented English). Results for in-domain LibriSpeech (native English) and CSJ (native Japanese) are also reported for reference: WER for English, CER for Japanese. The table compares different configurations of differentiable k-means (DiffKM) and multi-task learning (MTL; controlled by $\alpha$), together with two cluster centroid initializations: Init-L1 (trained with the listener’s L1) and Init-L2 (trained with the target language). The best results overall are in \textbf{bold}, and the best results within each initialization method are \underline{underlined}.}
	\label{tab:results}
    \resizebox{\textwidth}{!}{
    \begin{tabular}{lccc|ccc|c|c|ccc|c|c}
	  \toprule
          &   DiffKM  & MTL &$\alpha$ &  \multicolumn{5}{c|}{Init-L1 (Japanese)}
                & \multicolumn{5}{c}{Init-L2 (English)} \\\midrule
          &   &  & & \multicolumn{3}{c|}{ERJ}&LibriSpeech & CSJ & \multicolumn{3}{c|}{ERJ} &LibriSpeech & CSJ \\
          &    & &  &AE & JE\_all &JE\_w10 & test-\{clean, other\} & &AE & JE\_all &JE\_w10 & test-\{clean, other\} &  \\
        \midrule
	Baseline&$\times$  & $\times$& - & 13.3 & 52.7 & 66.7& 3.3/8.3 & -  & 12.7 & 54.4 & 68.9 & 3.3/8.2 & - \\  
 & $\checkmark$ & $\times$ &(0.0) & \underline{11.6} & 53.3 & 68.4&\underline{\textbf{2.9}/7.6} & -   & \underline{\textbf{11.4}} & \underline{52.6} & \underline{67.3} & 3.1/\textbf{7.4} & -\\ 
 \textbf{Ours}&$\checkmark$ & $\checkmark$ &0.3   & 11.8&\underline{\textbf{51.2}} & 66.0 & 3.0/7.6 & 10.5  & 11.7 & 53.9 & 69.1& \underline{3.0/\textbf{7.4}} & 10.8 \\ 
  & $\checkmark$ & $\checkmark$ &0.5   & 11.9&51.5 & \underline{\textbf{65.9}} & 3.1/7.6 & 10.3  & 12.0 & 54.5 & 70.4& 3.0/7.5 & \underline{10.2} \\ 
 & $\checkmark$ & $\checkmark$ &0.7  & 12.4&52.2 & 67.5 & 3.2/8.1 & \underline{\textbf{10.1}}   & 12.1 & 56.5 & 71.2& 3.2/7.7 & \underline{10.2} \\ 
	  \bottomrule
	\end{tabular}
    }
    \vspace*{-2.8mm}
\end{table*}
\subsection{ISIB reproduction in phonetic token-based ASR}
\label{subsec:isib}
In \cite{onda25_interspeech}, it was demonstrated that ISIB was reproduced in ASR using phonetic tokens obtained from HuBERT \cite{hubert} as intermediate representations. Experiments across multiple accents confirmed that conducting k-means clustering on a speaker’s native language achieved the highest recognition accuracy when recognizing foreign-accented speech, even outperforming the use of native speech from the target language spoken. This represents a pilot study toward accent-robust ASR using only native speech, serving as a technical implementation of ISIB, in which listeners sharing the native language with the speaker have an advantage in speech perception.

The framework proposed in \cite{onda25_interspeech} trains discrete tokens on the speaker’s native language (L1) and then uses them for downstream ASR training on the target language (L2). 
However, as discussed in Introduction, this approach is somewhat naive in that it relies solely on L1 to simulate non-native listeners’ speech perception.
Motivated by this, this study aims to more properly reproduce ISIB and improve ASR performance by optimizing tokens on both L1 and L2.

\subsection{Differentiable k-means for joint optimization of phonetic token-based ASR}
\label{subsec:diffkm}
In \cite{onda25c_interspeech}, a method was proposed that introduces differentiable k-means into phonetic token-based ASR, enabling the simultaneous optimization of token learning and downstream ASR training, thereby improving accuracy. 
The entire module was optimized for the following ASR loss.
\begin{align}
\label{eq:joint_loss}
\mathcal{L}
&=
\mathcal{L}^{\text{asr}}\Bigl(
    Y,\,
    \mathrm{ASR}\bigl(
        \mathrm{DiffKM}\bigl(\mathrm{SSL}(X;\,\theta_{\text{ssl}});\,M\bigr);\,
        \theta_{\text{asr}}
    \bigr)
\Bigr)
\end{align}
The components are: 1) the SSL model $\theta_{\text{ssl}}$, which performs feature extraction  $\mathrm{SSL}(X;\theta_{\text{ssl}})$ from the input speech $X$; 2) the stack of cluster centroids $M$, which is used for the differentiable k-means $\mathrm{DiffKM}(\cdot; M)$ that discretizes the SSL features; and 3) the ASR model $\theta_{\text{asr}}$, which predicts the target transcription $Y$ via $\mathrm{ASR}(\cdot;\theta_{\text{asr}})$.

As discussed in \cite{onda25c_interspeech}, this approach not only enhanced ASR performance but also changed the properties of the tokens, making them closer to phonetic representations that retain purer linguistic information. In this study, we extend this approach by introducing a multi-task learning framework and optimizing the tokens for the ASRs of both the speaker’s L1 and L2. This aims to more realistically reproduce the speech perception by non-native listeners and further improve ASR performance.

\subsection{Proposed method: multi-task learning with differentiable k-means for enhanced ISIB reproduction}
\label{subsec:proposed}
\begin{figure}[t]
  \centering
  \includegraphics[width=0.95\linewidth]{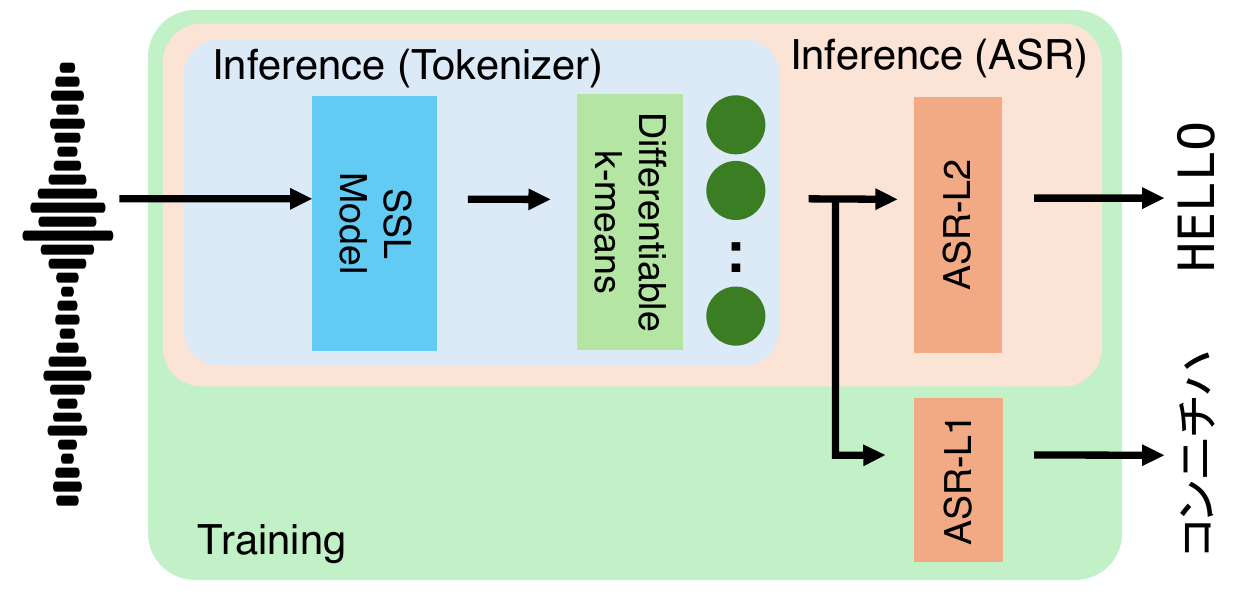}
   \vspace*{-2mm}
  \caption{Multi-task learning for advanced modeling of interlanguage speech intelligibility benefit. At inference time, the trained model can be used in two ways: as an ASR model or as a tokenizer.}
  \vspace*{-4mm}
  \label{fig:model}
\end{figure}
In this study, we combine the ideas presented in Sections \ref{subsec:isib} and \ref{subsec:diffkm} and propose a method for an advanced modeling of ISIB. This method optimizes tokens in a multi-task framework for ASR on both the speaker’s L1 and L2, using differentiable k-means.
In this method, the loss defined in Eq. (\ref{eq:joint_loss}) is extended as follows.
\begin{equation}
\label{eq:joint_loss_l1l2}
\resizebox{\columnwidth}{!}{$
\begin{aligned}
\mathcal{L}
&=
(1-\alpha)\, \mathcal{L}^{\text{asr-l2}}\Bigl(
    Y_{L2},\,
    \mathrm{ASR_{L2}}\bigl(
        \mathrm{DiffKM}\bigl(\mathrm{SSL}(X_{L2};\,\theta_{\text{ssl}});\,M\bigr);\,
        \theta_{\text{asr-l2}}
    \bigr)
\Bigr)
\\
&\quad + \alpha\, \mathcal{L}^{\text{asr-l1}} \Bigl(
    Y_{L1},\,
    \mathrm{ASR_{L1}}\bigl(
        \mathrm{DiffKM}\bigl(\mathrm{SSL}(X_{L1};\,\theta_{\text{ssl}});\,M\bigr);\,
        \theta_{\text{asr-l1}}
    \bigr)
\Bigr)
\end{aligned}
$}
\end{equation}
Fig. \ref{fig:model} illustrates the overview of the proposed method. During training, two ASR models, $\theta_{\text{asr-l1}}$ for L1 and $\theta_{\text{asr-l2}}$ for L2, are jointly learned in a multi-task manner. Both of the ASR models take tokens obtained through the operation of $\mathrm{DiffKM}(\mathrm{SSL}(\cdot;\theta_{\text{ssl}}); M)$ as a shared input. Through this process, the SSL model $\theta_{\text{ssl}}$  and the cluster centroids $M$ are optimized for both ASR tasks, and the fine-tuned tokens are expected to capture the phonetic characteristics of both L1 and L2, thereby enabling a more accurate modeling of ISIB. 
For each training batch, the ASR losses for L1 ($\mathcal{L}^{\text{asr-l1}}$) and L2 ($\mathcal{L}^{\text{asr-l2}}$) are computed separately, with the results for $\mathrm{ASR_{L1}}(\cdot;\theta_{\text{asr-l1}})$ and $\mathrm{ASR_{L2}}(\cdot;\theta_{\text{asr-l2}})$, respectively. They are then combined with a weighting factor $\alpha$ and used to update the entire model parameters.

At inference time, this model can be used both 1) as an ASR model and 2) as a tokenizer. These two usages correspond to the two experimental scenarios described in Sec. \ref{subsec:scenario}, which are based on the availability of accented speech.
When used as an ASR model (as shown in the ``Inference (ASR)" of Fig. \ref{fig:model}), we directly use the trained ASR models for L2 $\theta_{\text{asr-l2}}$ to recognize foreign-accented speech. 
On the other hand, when used as a tokenizer (as shown in the ``Inference (Tokenizer)” of Fig. \ref{fig:model}), only the SSL model $\theta_{\text{ssl}}$ and the differentiable k-means with the trained cluster centroids $M$ are used to extract tokens for further applications. 
In this case, the model, which is trained only with native speech data, is used to obtain tokens that capture phonetic characteristics of both L1 and L2.

\section{Experiments}
\label{sec:exp}
\subsection{Experimental setup}
\label{subsec:setup}
In this study, we focus on the case of Japanese-accented English, where the speaker’s L1 is Japanese and the L2 is English. Although experiments on other accents are omitted due to space limitations, our prior pilot study across multiple accents \cite{onda25_interspeech} showed consistent trends, suggesting that the proposed framework could be generalizable beyond Japanese-accented English. Model training was conducted using ESPnet \cite{watanabe18_interspeech}. As the SSL model, we employed the last layer of HuBERT-base\footnote{\url{https://hf.co/facebook/hubert-base-ls960}}, which was reported in \cite{onda25_interspeech} to achieve the highest accuracy. The number of clusters was set to 2000, and the configuration for differentiable k-means followed \cite{onda25c_interspeech}. For the ASR models, we used a joint CTC/attantion-based encoder-decoder (AED) model \cite{ctcaed} for L2 (English) with a CTC weight of 0.3, while for L1 (Japanese), we used a CTC-only model. For training data, we used LibriSpeech-960h \cite{libri} for English and CSJ (661h) \cite{maekawa03_sspr} for Japanese. Both corpora consist exclusively of native speech in each language. As output text tokens, we used 5,000 BPE tokens for English and a character-based \textit{katakana} representation for Japanese. \textit{Katakana} represents sounds rather than meanings (as \textit{kanji}), enabling tokens to better capture phonetic characteristics of Japanese, which influence the English pronunciation by Japanese speakers.

Following \cite{onda25c_interspeech}, training was conducted in two stages. In the first stage, the SSL model $\theta_{\text{ssl}}$ (HuBERT-base) and the cluster centroids $M$, which were initialized using standard k-means, were frozen, and only the ASR models $\theta_{\text{asr-l1}}$ and $\theta_{\text{asr-l2}}$ were trained for 20 epochs with a learning rate of 1e-3. In the second stage, the entire module including $\theta_{\text{ssl}}$ and $M$ was fine-tuned jointly for 20 epochs with a learning rate of 1e-5.
The weight $\alpha$ for the L1 ASR loss (in Eq.(\ref{eq:joint_loss_l1l2})), used for multi-task learning, was varied across three settings: 0.3, 0.5, and 0.7. 
We consider two baselines. The first uses discrete tokens obtained with standard non-differentiable k-means without any updating, following the experimental setup of \cite{onda25_interspeech}. The second corresponds to $\alpha = 0$, where the model is optimized solely for English ASR; in this case, the model architecture is equivalent to that of \cite{onda25c_interspeech}.
When the tokens were updated, the cluster centroids $M$ were initialized using those learned by standard k-means. We compared two initialization strategies: centroids trained on L1 (Japanese) data and those trained on L2 (English) data. For this initialization, we used JVS \cite{takamichi2019jvs} for L1 and a 30-hour random subset of LibriSpeech train-clean-100 for L2 to train k-means clusters, following \cite{onda25_interspeech}.

\subsection{Two experimental scenarios: native-only \& accent-adapted}
\label{subsec:scenario}
As mentioned in \ref{subsec:proposed}, this study explores two experimental scenarios 
based on the availability of the target non-native speech, corresponding to the
two usages (ASR and tokenizer) of our model.

\noindent \textbf{Native-only: }if no accented speech is available, rather than the  native speech used for model training, this model can be used as an ASR system. 
This means that the ASR model component for L2 $\theta_{\text{asr-l2}}$ recognize out-of-domain non-native speech directly.

\noindent \textbf{Accent-adapted: }If a small amount of target non-native speech is available, the model can be used as a tokenizer, and the obtained tokens can then be used to train a separate accent-specific, discrete token-based ASR model. 

\subsection{ERJ corpus}
\label{subsec:erj}
For evaluation, we used the ERJ corpus \cite{Minematsu2004DevelopmentOE}, which consists of speech data from Japanese learners of English and native American English speakers reading the same set of sentences and words. In this study, we employed the phonetic sentence section as the evaluation data, which contains 460 phonetically balanced sentences. 

In Sec. \ref{subsec:asr}, where we report the results for \textbf{native-only} scenario, the accented speech data from ERJ are used as out-of-domain test set. We assess the performance of the models trained only with native speech when recognizing non-native speech.
For this evaluation, we also used the pronunciation scores included in ERJ, which were annotated by native raters. We adopt the segmental score as a metric of accent strength for detailed analyses.
We evaluated ASR performance both on the overall set of Japanese learners’ speech (JE\_all) and separately on the worst 10 speakers with the lowest segmental scores (JE\_w10), in order to investigate the impact of the degree of accents. We also evaluated recognition performance on native English by American speakers (AE) as a reference.

In Sec. \ref{subsec:asr_2}, for the results of \textbf{accent-adapted} scenario, the ERJ data is first converted into tokens with the same model used as a tokenizer, and then we used them to train a separate discrete token-based ASR model specialized for Japanese-accented English.

\subsection{ASR performance with the native-only scenario}
\label{subsec:asr}
In this section, we evaluate the effectiveness of our method in the \textbf{native-only} scenario.
Table \ref{tab:results} presents recognition results for out-of-domain Japanese-accented English (ERJ-JE\_all and JE\_w10) and native American English (ERJ-AE). For reference, the accuracy on in-domain native L2 (LibriSpeech) and native L1 (CSJ) is also shown. Word error rate (WER) was used to evaluate English recognition performance, while character error rate (CER) was used for Japanese.

In the baseline using vanilla k-means (DiffKM: $\times$, MTL: $\times$), tokens trained on L1 resulted in higher accuracy for JE than those trained on L2, reproducing the findings of previous studies that ISIB occurs in discrete token-based ASR \cite{onda25_interspeech}. 
Next, we examine the effect of differentiable k-means and multi-task learning on ASR performance. 
When initialized with Japanese (Init-L1), optimizing tokens solely for English (DiffKM: $\checkmark$, MTL: $\times$) improved accuracy for native English (both in-domain and out-of-domain) but degraded performance on Japanese-accented English. 
Since only native speech is used for training in our setting, such a trade-off caused by a domain mismatch is expected. 
However, introducing multi-task learning (DiffKM: $\checkmark$, MTL: $\checkmark$) led to improvements: for the overall Japanese-accented English (JE\_all), the highest accuracy was achieved at $\alpha = 0.3$, while for the strongly accented JE\_10 subset, the best performance occurred at $\alpha = 0.5$. Further increasing $\alpha$ resulted in a decline in accuracy. 
These results demonstrate the distinct roles of the two components: differentiable k-means enables discrete tokens to be adapted to arbitrary objectives, while multi-task learning allows the L2 ASR to capture phonetic characteristics of L1, thereby enhancing ASR performance on out-of-domain accented speech.
It can be assumed that strongly accented speech is closer to the phonetic characteristics of L1, which explains why the optimal $\alpha$ tends to be higher. This can be interpreted as the result of simulating non-native listeners' speech perception by appropriately fine-tuning tokens, which were initially trained on L1, and then optimized for L2 ASR as well. 
However, different results were observed when the tokens were initialized using L2 data (Init-L2). In this case, the highest recognition accuracy was observed when optimized solely for English ($\alpha = 0$), for both native English and Japanese-accented English. This is likely because initializing with L2 caused the model to focus more on the English-like features in Japanese-accented English, which contains both Japanese-like and English-like properties. As a result, deviations from native English due to Japanese accent were treated as noise, and the model is thought to have learned to suppress them during training. 
For in-domain native Japanese, recognition accuracy consistently improved with larger $\alpha$, regardless of the initialization setting.

Overall, the results showed that for native English, the highest accuracy was achieved when tokens were initialized with English and fine-tuned solely for English ASR ($\alpha = 0$), whereas for Japanese-accented English, the best performance was obtained when tokens were initialized with Japanese and fine-tuned in a multitask setting for both L1 and L2 ($\alpha = 0.3, 0.5$). This can be interpreted as a more advanced reproduction of ISIB, where native English is easier for native speakers to recognize, while Japanese-accented English is easier for Japanese learners of English.

\subsection{ASR performance with the accent-adapted scenario}
\label{subsec:asr_2}
Next, we verify the effectiveness in the \textbf{accent-adapted} scenario, using a limited amount of accented speech for further training. First, the Japanese-accented English speech in ERJ was randomly divided into training, validation, and test sets at a ratio of 8:1:1 in a speaker-independent manner. This means that no speaker appeared in more than one set. The training set contained 11.2 hours of speech; however, to simulate scenarios with more limited amount of data, we also randomly extracted 2-hour and 5-hour subsets from the training data and compared the performance of ASR models trained on these subsets. Here, the same models as used in Sec. \ref{subsec:asr}, which are trained only on native speech, were used to convert the accented speech in ERJ into tokens (see ``Inference (Tokenizer)" part of Fig. \ref{fig:model}). The tokens were then used to train another discrete token-based ASR model. CTC/AED models were trained with a learning rate of 5e-3 until convergence. Training was conducted for 60 epochs for the full 11.2h ERJ dataset, 100 epochs for the 5h subset, and 250 epochs for the 2h subset.

Table \ref{tab:results_accent} shows the recognition performance on the ERJ test set for models trained with different amounts of data. When using the full training data, the baseline standard k-means trained with L1 achieved the highest performance. However, in scenarios with limited training data (2h and 5h), the proposed method initialized with L1 and using $\alpha = 0.3$ achieved the best performance. In the native-only scenario shown in Table \ref{tab:results}, that setting also achieved the highest performane, but the improvement was marginal. In this accent-adapted scenario, however, the proposed method demonstrated a substantial improvement. A relative reduction in WER of 19.3\% was observed for the 2h subset, and 19.4\% for 5h, compared to the vanilla k-means baseline of Init-L1. This suggests that the tokens fine-tuned via multi-task learning on both L1 and L2 ASR acquire representations that enable efficient learning of accented speech recognition even from limited training data.

\begin{table}[tb]
	\centering
    \vspace*{-2.2mm}
	\caption{Recognition accuracy (WER) on the in-domain test split (speaker-independent) when a discrete token-based ASR was trained on the limited Japanese-accented speech from ERJ, with tokenizers derived from models trained on native LibriSpeech and CSJ. L1/L2 denote the language used for k-means centroid initialization.}
	\label{tab:results_accent}
    \resizebox{\linewidth}{!}{
    \begin{tabular}{lccc|cc|cc|cc}
	  \toprule
               & DiffKM&MTL & $\alpha$ & \multicolumn{2}{c|}{2h} &  \multicolumn{2}{c|}{5h} &  \multicolumn{2}{c}{all (11.2h)}\\
             &  &  &  & L1 & L2 &   L1 & L2&   L1 & L2\\\midrule

	Baseline  &$\times$  & $\times$ & -&43.0& 43.1& 28.8& 29.2& \textbf{\phantom{1}8.0}&\phantom{1}8.7 \\  
   &$\checkmark$  & $\times$&(0.0) &39.8 & 41.0& 23.8& 25.6& 11.0& 10.7\\  
 \textbf{Ours}  & $\checkmark$  & $\checkmark$ & 0.3&\textbf{34.7}& 36.5& \textbf{23.2}& 23.9& \phantom{1}9.5&\phantom{1}9.8 \\  
   & $\checkmark$  & $\checkmark$ &0.5&34.8 & 36.4& 28.3 &24.6& \phantom{1}8.1& 10.7\\  
   & $\checkmark$  & $\checkmark$& 0.7&36.1& 39.8& 25.9& 26.6& \phantom{1}8.6& 13.3\\  
	  \bottomrule
	\end{tabular}

    }
    \vspace*{-4mm}
\end{table}

\section{Conclusions}
In this study, we proposed a method to enhance the robustness of discrete token-based ASR to foreign-accented speech through an advanced modeling of interlanguage speech intelligibility benefit (ISIB). By using differentiable k-means to optimize the entire module via multi-task learning of L1 and L2 ASR, the method addresses a limitation of conventional approaches, in which tokens are trained solely on L1. This enables modeling of speech perception based on both L1 and L2, similar to human non-native listeners, and successfully improves recognition accuracy for foreign-accented speech.

Our method is designed to effectively leverage abundant native speech data available and was validated under two scenarios: 1) native-only (in Sec. \ref{subsec:asr}) and 2) accent-adapted (in Sec. \ref{subsec:asr_2}). In the former scenario, the model trained only with native speech was used as an ASR system for inference, and its recognition performance was evaluated on out-of-domain foreign-accented speech. In the latter scenario, the same model served as a tokenizer to train a separate discrete token-based ASR model on a limited amount of additional foreign-accented speech, and its in-domain performance was evaluated. In both scenarios, our method outperformed the baselines, and especially in the latter scenario, the method achieved an approximately 20\% relative improvement. This indicates that the proposed method can effectively leverage speakers’ L1 native speech to enhance the accent-robustness of ASR.

Future work includes evaluating the effectiveness of the proposed approach for accents other than Japanese-accented English, and developing models capable of handling scenarios in which the speaker’s native language is unknown.

\section{Acknowledgements}
This work was supported by AIST policy-based budget project ``R\&D on Generative AI Foundation Models for the Physical Domain."
\bibliographystyle{IEEEbib}
\bibliography{references}

\end{document}